\documentclass[11pt]{article}

\usepackage[cmtip,arrow]{xy}
\usepackage{pb-diagram,pb-xy}

\newlength{\vshift}
\newlength{\hshift}
\usepackage{amssymb,amsopn}

\def\nn{\nonumber }
\def\la{\lambda}

\def\be{\beta}
\def\a{\alpha}

\def\g{\gamma}

\def\ds{\stackrel{\star}{,}}

\def\p{\partial}

\def\p{\partial}

\def\Str{{\rm STr}}
\def\tr{{\rm Tr}}
\def\slash{{\rlap /}}

\def\nn{\nonumber}
\def\be{\begin{equation}}             \def\ee{\end{equation}}
\def\ba#1{\begin{array}{#1}}          \def\ea{\end{array}}
\def\bea{\begin{eqnarray} }           \def\eea{\end{eqnarray} }
\def\beann{\begin{eqnarray*} }        \def\eeann{\end{eqnarray*} }
\def\beal{\begin{eqalign}}            \def\eeal{\end{eqalign}}
             
\def\bsubeq{\begin{subequations}}     \def\esubeq{\end{subequations}}
\def\bitem{\begin{itemize}}           \def\eitem{\end{itemize}}
                    
\def\pa{\partial}

\begin{document}

\begin{titlepage}

$\,$

\vspace{1.5cm}
\begin{center}

{\LARGE{\bf D-deformed Wess-Zumino model and its renormalizability properties}}

\vspace*{1.3cm}

{{\bf Marija Dimitrijevi\' c and
Voja Radovanovi\' c}}

\vspace*{1cm}

University of Belgrade, Faculty of Physics\\
Studentski trg 12, 11000 Belgrade, Serbia \\[1em]


\end{center}

\vspace*{2cm}

\begin{abstract}

Using the methods developed in earlier papers we analyze a new
type of deformation of the superspace. The twist we use to deform
the $N=1$ SUSY Hopf algebra is non-hermitian and is given in terms
of the covariant derivatives $D_\alpha$. A SUSY invariant
deformation of the Wess-Zumino action is constructed and compared
with results already known in the literature. Finally, by
calculating divergences of the two-point Green functions a
preliminary analysis of renormalizability properties of the
constructed model is done. As expected, there is no
renormalization of mass and no tadpole diagrams appear.

\end{abstract}
\vspace*{1cm}



\vspace*{1cm}
\quad\scriptsize{eMail:
dmarija,rvoja@phy.bg.ac.yu}
\vfill

\end{titlepage}\vskip.2cm

\newpage
\setcounter{page}{1}
\newcommand{\Section}[1]{\setcounter{equation}{0}\section{#1}}
\renewcommand{\theequation}{\arabic{section}.\arabic{equation}}

\section{Introduction}

The idea of noncommuting spacetime coordinates goes back to
Heisenberg who suggested \cite{Heis} that uncertainty relations
between coordinates could resolve the ultraviolet (UV) divergences
arising in quantum field theories. The issue was then investigated
by Snyder in \cite{Snyder}, but did not attract much interest at
the time. However, in the last two decades noncommutative geometry
has found applications in many branches of physics such as quantum
field theory and particle physics, solid state physics and many
others. Comprehensive reviews on the subject can be found in
references \cite{Connes}, \cite{Landi}, \cite{madorebook},
\cite{Castellani1}, \cite{revDN},\cite{revSz1}, \cite{revSz2}.

Having in mind problems which physics encounters at small scales
(high energies), in recent years attempts were made to combine
supersymmetry (SUSY) with noncommutative geometry. Different
models were constructed, see for example \cite{luksusy},
\cite{MWsusy}, \cite{nonantisusy}, \cite{Seiberg}, \cite{Ferrara},
\cite{14}. Some of these models emerge naturally as low energy
limits of string theories in backgrounds with a constant
Neveu-Schwarz two form and/or a constant Ramond-Ramond two form.
In \cite{Seiberg} the anticommutation relations between the
fermionic coordinates were modified in the following way
\begin{equation}
\{ \theta^\alpha \ds \theta^\beta\} =C^{\alpha\beta} , \quad \{
\bar{\theta}_{\dot\alpha} \ds \bar{\theta}_{\dot\beta}\} = \{
\theta^\alpha \ds \bar{\theta}_{\dot\alpha}\} = 0\ ,
\label{eucliddef}
\end{equation}
where $C^{\alpha\beta} = C^{\beta\alpha}$ is a complex constant
symmetric matrix. The analysis is done in Euclidean space, since
the deformation (\ref{eucliddef}) is hermitian only in Euclidean
signature where undotted and dotted spinors are not related by the
usual complex conjugation. Note that the $\star$-product used in
(\ref{eucliddef}) is also well defined in Minkowskian signature
although in that case it is not hermitian \cite{Ferrara}.  The
chiral coordinates $y^m=x^m+i\theta\sigma^m\bar\theta$ commute in
this setting, therefore the notion of chirality is preserved, i.e.
the $\star$-product of two chiral superfields is again a chiral
superfield. On the other hand, the constructed models break one
half of the $N=1$ SUSY so they are invariant only under the
so-called $N=1/2$ SUSY. Renormalizability of the Wess-Zumino
models with this deformation was considered, see for example
\cite{1/2WZrenorm}. Some of the obtained results are: the
renormalizability is lost already at the one loop level, but it
can be restored by adding to the classical action new couplings
(interaction terms) which depend on the deformation parameter.
Also, a pure gauge sector which is supergauge invariant and one
loop renormalizable can be constructed \cite{1/2gauge}. Recently,
a renormalizable $N=1/2$ super-Yang-Mills model with interacting
matter was constructed in \cite{1/2SYMmatter}.

Another type of deformation was introduced in \cite{Ferrara}.
There the product of two chiral superfields is not a chiral
superfield but the model is invariant under the full
supersymmetry. A deformation of the Hopf algebra of SUSY
transformations by a twist was considered in \cite{chiralstar}.

In our previous paper \cite{miSUSY} we applied the twist formalism
to deform the Hopf algebra of $N=1$ SUSY transformations. Our
choice of the twist is different from that in \cite{chiralstar}.
We work in Minkowski space-time and choose a hermitian twist. As
undotted and dotted spinors are related by the usual complex
conjugation, we obtain
\begin{equation}
\{ \theta^\alpha \ds \theta^\beta\} = C^{\alpha\beta} , \quad \{
\bar{\theta}_{\dot\alpha} \ds \bar{\theta}_{\dot\beta}\} =
\bar{C}_{\dot{\alpha}\dot{\beta}}, \quad \{ \theta^\alpha \ds
\bar{\theta}_{\dot\alpha}\} = 0\ , \label{minkdef}
\end{equation}
with $\bar{C}_{\dot{\alpha}\dot{\beta}} = (C_{\alpha\beta})^*$.
The deformed Wess-Zumino Lagrangian was formulated and analyzed;
the action which follows is invariant under the twisted SUSY
transformations. Superfields transform in the undeformed way,
while the Leibniz rule for SUSY transformations (when they act on
a product of superfields) is modified. Since the action is
non-local it is difficult (but not impossible) to discuss
renormalizability properties of the model.

Nevertheless, we are interested in renormalizability properties of
theories with twisted symmetries. It is important to understand
whether deforming (by a twist) of symmetries spoils some of the
renormalizability properties of SUSY invariant theories.
Therefore, in this paper we analyze a simpler model with the twist
given by
\begin{equation}
{\cal F} = e^{\frac{1}{2}C^{\alpha\beta}D_\alpha \otimes D_\beta
} ,\label{intro-twist}
\end{equation}
where $C^{\alpha\beta} = C^{\beta\alpha}\in \mathbb{C}$ is a
complex constant matrix and $D_\alpha = \partial_\alpha -
i\sigma_m^{\
\alpha\dot{\alpha}}\bar{\theta}_{\dot{\alpha}}\partial_m$ are the
SUSY covariant derivatives.

Following the method of \cite{miSUSY} in the next section we
introduce the deformation and the $\star$-product which follows
from it. Due to our choice of the twist (\ref{intro-twist}), the
coproduct of SUSY transformations\footnote{We only consider the
$N=1$ SUSY in this paper. The generalization to $N=2$ SUSY and
higher can be obtained by following the same steps as in Section
2.} remains undeformed, leading to the undeformed Leibniz rule.
Being interested in deformations of the Wess-Zumino model, we
discuss chiral fields and their products. The product of two
chiral fields is not a chiral field and we have to use the
projectors defined in \cite{wessbook} to separate chiral and
antichiral parts. All possible invariants are listed in Section 4
and the deformed Wess-Zumino action is constructed in Section 5.
Using the background field method we then analyze two-point
functions and their divergences. Finally, we give some comments
and compare our results with the results already present in the
literature. Some details of the calculations are collected in
appendices A and B.

\section{$D$-deformation of the Hopf algebra of SUSY transformations}

The undeformed superspace is generated by the coordinates $x$,
$\theta$ and $\bar{\theta}$ which fulfill
\begin{eqnarray}
\lbrack x^m, x^n \rbrack &=& \lbrack x^m,
\theta^\alpha \rbrack = \lbrack x^m, \bar{\theta}_{\dot\alpha} \rbrack = 0 ,\nonumber\\
\{ \theta^\alpha , \theta^\beta\} &=& \{ \bar{\theta}_{\dot\alpha}
, \bar{\theta}_{\dot\beta}\} = \{ \theta^\alpha ,
\bar{\theta}_{\dot\alpha}\} = 0 , \label{undefsupsp}
\end{eqnarray}
with $m=0,\dots 3$ and $\alpha, \beta =1,2$. To $x^m$ we refer as
to bosonic and to $\theta^\alpha$ and $\bar{\theta}_{\dot\alpha}$
we refer as to fermionic coordinates. Also, $x^2= x^m x_m =
-(x^0)^2+(x^1)^2 + (x^2)^2 + (x^3)^2 $. A general superfield
$F(x,\theta, \bar{\theta})$ can be expanded in powers of $\theta$
and $\bar{\theta}$
\begin{eqnarray}
F(x, \theta, \bar{\theta}) &=&\hspace*{-2mm} f(x) + \theta\phi(x)
+ \bar{\theta}\bar{\chi}(x)
+ \theta\theta m(x) + \bar{\theta}\bar{\theta} n(x) + \theta\sigma^m\bar{\theta}v_m\nonumber\\
&& + \theta\theta\bar{\theta}\bar{\lambda}(x) + \bar{\theta}\bar{\theta}\theta\varphi(x)
+ \theta\theta\bar{\theta}\bar{\theta} d(x) .\label{F}
\end{eqnarray}
Under the infinitesimal SUSY transformations it transforms in the following way
\begin{equation}
\delta_\xi F = \big(\xi Q + \bar{\xi}\bar{Q} \big) F, \label{susytr}
\end{equation}
where $\xi$ and $\bar{\xi}$ are constant anticommuting parameters
and $Q$ and $\bar{Q}$ are the SUSY generators
\begin{eqnarray}
Q_\alpha &=& \p_\alpha
- i\sigma^m_{\ \alpha\dot{\alpha}}\bar{\theta}^{\dot{\alpha}}\p_m, \label{q}\\
\bar{Q}^{\dot{\alpha}} &=& \bar{\p}^{\dot{\alpha}} -
i\theta^\alpha \sigma^m_{\
\alpha\dot{\beta}}\varepsilon^{\dot{\beta}\dot{\alpha}}\p_m
.\label{barq}
\end{eqnarray}

As in \cite{defgt}, \cite{miSUSY}, we introduce a deformation of
the Hopf algebra of infinitesimal SUSY transformations by choosing
the twist $\cal{F}$ in the following way
\begin{equation}
{\cal F} = e^{\frac{1}{2}C^{\alpha\beta}D_\alpha \otimes D_\beta
} ,\label{twist}
\end{equation}
with the complex constant matrix $C^{\alpha\beta} =
C^{\beta\alpha}\in \mathbb{C}$. Note that this twist is not
hermitian, ${\cal F}^* \neq {\cal F}$. The usual complex
conjugation is denoted by "$*$". It can be shown \cite{twist2}
that (\ref{twist}) satisfies all the requirements for a twist
\cite{chpr}. The Hopf algebra of SUSY transformation does not
change since
\begin{equation}
\{ Q_\alpha , D_\beta\} = \{ \bar{Q}_{\dot{\alpha}} , D_\beta\} =
0 \label{komQD}
\end{equation}
and it is given by
\begin{itemize}
\item
algebra
\begin{eqnarray}
&& \{ Q_\alpha , Q_\beta\} = \{ \bar{Q}_{\dot{\alpha}} , \bar{Q}_{\dot{\beta}}\} = 0, \quad
\{ Q_\alpha , \bar{Q}_{\dot{\beta}} \} = 2i\sigma^m_{\ \alpha\dot{\beta}}\p_m ,\nonumber\\
&& \hspace*{1cm}[\p_m, \p_n] = [\p_m, Q_\alpha] = [\p_m, \bar{Q}_{\dot{\alpha}}] =0 .\label{Qcom}
\end{eqnarray}

\item
coproduct
\begin{eqnarray}
&&\Delta Q_\alpha = Q_\alpha \otimes 1 + 1\otimes Q_\alpha, \quad
\Delta\bar{Q}_{\dot{\alpha}} = \bar{Q}_{\dot{\alpha}} \otimes 1 + 1\otimes \bar{Q}_{\dot{\alpha}}, \nonumber \\
&&\hspace*{2cm} \Delta \p_m = \p_m \otimes 1 + 1\otimes \p_m. \label{coprQ}
\end{eqnarray}

\item
counit and antipode
\begin{eqnarray}
&& \varepsilon(Q_\alpha)=\varepsilon(\bar{Q}_{\dot{\alpha}})=
\varepsilon(\p_m) =0, \nonumber \\
&& S(Q_\alpha) = -Q_\alpha, \quad S(\bar{Q}_{\dot{\alpha}}) =
-\bar{Q}_{\dot{\alpha}}, \quad S(\p_m) = -\p_m
.\label{undefcounitantipod}
\end{eqnarray}

\end{itemize}
This means that the full supersymmetry is preserved.

Strictly speaking, the twist (\ref{twist}) does not belong to the
universal enveloping algebra of the Lie algebra of infinitesimal
SUSY transformations. Therefore, to be mathematically correct we
should enlarge the algebra (\ref{Qcom}) by introducing the
relations for the operators $D_{\alpha}$ as well. Note that the
same happened in \cite{miSUSY}, where the twist was given by
\begin{equation}
{\cal F} = e^{\frac{1}{2}C^{\alpha\beta}\p_\alpha \otimes\p_\beta
+\frac{1}{2}\bar{C}_{\dot{\alpha}\dot{\beta}}\bar{\p}^{\dot{\alpha}}
\otimes\bar{\p}^{\dot{\beta}} } ,\label{oldtwist}
\end{equation}
with the complex constant matrix $C^{\alpha\beta} =
C^{\beta\alpha}$ and $C^{\alpha\beta}$ and
$\bar{C}^{\dot{\alpha}\dot{\beta}}$ were related by the usual
complex conjugation. There we had to enlarge the algebra by adding
the relations for the fermionic derivatives $\p_\alpha$ and
$\bar{\p}^{\dot{\alpha}}$.

The inverse of the twist (\ref{twist})
\begin{equation}
{\cal F}^{-1} = e^{-\frac{1}{2}C^{\alpha\beta}D_\alpha
\otimes D_\beta } ,\label{invtwist}
\end{equation}
defines the $\star$-product. For two arbitrary superfields $F$ and
$G$ the $\star$-product reads
\begin{eqnarray}
F\star G &=& \mu_\star \{ F\otimes G \} \nonumber\\
&=& \mu \{ {\cal F}^{-1}\, F\otimes G\} \nonumber\\
&=& \mu \{ e^{-\frac{1}{2}C^{\alpha\beta}D_\alpha \otimes D_\beta
} F\otimes G \} \nonumber \\
&=& F\cdot G - \frac{1}{2}(-1)^{|F|}C^{\alpha\beta}(D_\alpha
F)\cdot(D_\beta G) \nonumber\\
&&- \frac{1}{8}C^{\alpha\beta}C^{\gamma\delta}(D_\alpha D_\gamma F)\cdot
(D_\beta D_\delta G)
, \label{star}
\end{eqnarray}
where $|F| = 1$ if $F$ is odd (fermionic) and $|F|=0$ if $F$ is
even (bosonic). The second line of (\ref{star}) is the definition
of the $\mu_\star$ multiplication. No higher powers of
$C^{\alpha\beta}$ appear since the derivatives $D_\alpha$ are
Grassmanian. The $\star$-product (\ref{star}) is
associative\footnote{The associativity of the $\star$-product
follows from the cocycle condition \cite{chpr} which the twist
${\cal F}$ has to fulfill
\begin{equation}
{\cal F}_{12}(\Delta\otimes id){\cal F} = {\cal F}_{23}(id\otimes\Delta){\cal F}, \label{cocycle}
\end{equation}
where ${\cal F}_{12} = {\cal F}\otimes 1$ and ${\cal F}_{23} =
1\otimes {\cal F}$. It can be shown that the twist (\ref{twist})
indeed fulfills this condition, see for details \cite{twist2}.},
noncommutative and in the zeroth order in the deformation
parameter $C_{\alpha\beta}$ it reduces to the usual pointwise
multiplication. One should also note that it is not hermitian,
\begin{equation}
(F\star G)^* \neq G^* \star F^* . \label{complconj}
\end{equation}

The $\star$-product (\ref{star}) leads to
\begin{eqnarray}
\{ \theta^\alpha \ds \theta^\beta \} &=& C^{\alpha\beta}, \quad \{
\bar{\theta}_{\dot\alpha}\ds \bar{\theta}_{\dot\beta}\} = \{ \theta^\alpha \ds
\bar{\theta}_{\dot\alpha} \} = 0 ,\nonumber \\
\lbrack x^m \ds x^n \rbrack &=& -C^{\alpha\beta}(\sigma^{mn}\varepsilon)_{\alpha\beta}\bar{\theta}\bar{\theta} , \nonumber \\
\lbrack x^m \ds \theta^\alpha \rbrack
&=& -iC^{\alpha\beta}\sigma^m_{\beta\dot{\beta}}\bar{\theta}^{\dot{\beta}},
\quad \lbrack x^m \ds \bar{\theta}_{\dot\alpha} \rbrack = 0 .
\label{thetastar}
\end{eqnarray}
The chiral coordinates $y^m$ also do not commute
\begin{eqnarray}
\lbrack y^m \ds y^n \rbrack &=&
-8\bar{\theta}\bar{\theta}C^{\alpha\beta}
(\sigma^{mn}\varepsilon)_{\alpha\beta}
. \label{ystar}
\end{eqnarray}
Other (anti)commutation relations follow in a similar way.

Relations (\ref{thetastar}) enable us to define the deformed
superspace. It is generated by the usual bosonic and fermionic
coordinates (\ref{undefsupsp}) while the deformation is contained
in the new product (\ref{star}). From (\ref{thetastar}) it follows
that both fermionic and bosonic part of the superspace are
deformed. This is different from \cite{miSUSY} where only the
fermionic coordinates were deformed.

The deformed infinitesimal SUSY transformation is defined as
\begin{eqnarray}
\delta^\star_\xi F &=& \big(\xi Q + \bar{\xi}\bar{Q} \big) F .\label{defsusytr}
\end{eqnarray}
Since the coproduct (\ref{coprQ}) is undeformed, the usual (undeformed) Leibniz rule follows. Then
the $\star$-product of two superfields is again a superfield. Its
transformation law is given by
\begin{eqnarray}
\delta^\star_\xi (F\star G) &=& \big(\xi Q + \bar{\xi}\bar{Q} \big) (F\star G) \nonumber\\
&=& (\delta^\star_\xi F)\star G + F\star (\delta^\star_\xi G) . \label{deftrlaw}
\end{eqnarray}

\section{Chiral fields}

Since we are interested in possible deformations of the
usual\footnote{In this paper ``usual`` always refers to
undeformed, that is to the case $C_{\alpha\beta}=0$.} Wess-Zumino
action, we now analyze chiral fields and their $\star$-products.

A chiral field $\Phi$ fulfills $\bar{D}_{\dot{\alpha}}\Phi =0$,
where $\bar{D}_{\dot{\alpha}} = -\bar{\p}_{\dot{\alpha}} -
i\theta^\alpha \sigma^m_{\ \alpha\dot{\alpha}}\p_m$ and
$\bar{D}_{\dot{\alpha}}$ is related to $D_\alpha$ by the usual
complex conjugation. In terms of the component fields the chiral
superfield $\Phi$ is given by
\begin{eqnarray}
\Phi(x, \theta, \bar{\theta}) &=& A(x) + \sqrt{2}\theta^\alpha\psi_\alpha(x)
+ \theta\theta H(x) + i\theta\sigma^l\bar{\theta}(\p_l A(x)) \nonumber\\
&& -\frac{i}{\sqrt{2}}\theta\theta(\p_m\psi^\alpha(x))\sigma^m_{\ \alpha\dot{\alpha}}\bar{\theta}^{\dot{\alpha}}
+ \frac{1}{4}\theta\theta\bar{\theta}\bar{\theta}(\Box A(x)).\label{chiral}
\end{eqnarray}
The $\star$-product of two chiral fields reads
\begin{eqnarray}
\Phi\star\Phi &=& \Phi\cdot\Phi-\frac{1}{8}C^{\alpha\beta}C^{\gamma\delta}D_\alpha
D_\gamma\Phi D_\beta D_\delta \Phi \nonumber\\
&=& \Phi\cdot\Phi -\frac{1}{32}C^2 (D^2\Phi)(D^2\Phi)\nonumber\\
&=& A^2 - \frac{C^2}{2}H^2 + 2\sqrt{2}A\theta^\alpha\psi_\alpha \nonumber\\
&& -i\sqrt{2}C^2 H\bar{\theta}_{\dot{\alpha}}
\bar{\sigma}^{m\dot{\alpha}\alpha}(\p_m\psi_\alpha)
+ \theta\theta \Big( 2AH - \psi\psi\Big) \nonumber\\
&& + C^2\bar{\theta}\bar{\theta}\Big( - H\Box A
+ \frac{1}{2}(\p_m\psi)\sigma^m\bar{\sigma}^l(\p_l\psi) \big) \Big) \nonumber\\
&& + i\theta\sigma^m\bar{\theta}\Big( \p_m (A^2) +C^2H\p_m H \Big) \nonumber\\
&& +
i\sqrt{2}\theta\theta\bar{\theta}_{\dot{\alpha}}\bar{\sigma}^{m\dot{\alpha}\alpha}
\big( \p_m(\psi_\alpha A)\big)\nonumber\\
&&+\frac{\sqrt{2}}{2}\bar\theta\bar\theta C^2(-H\theta\Box  \psi+\theta\sigma^m\bar\sigma^n\p_n \psi\p_m H)\nonumber\\
&&+ \frac{1}{4}\theta\theta\bar{\theta}\bar{\theta} (\Box
A^2-\frac12C^2\Box H^2) , \label{phistarphi}
\end{eqnarray}
where $C^2 =
C^{\alpha\beta}C^{\gamma\delta}\varepsilon_{\alpha\gamma}\varepsilon_{\beta\delta}$.
Because of the $\bar{\theta}$, $\bar{\theta}\bar{\theta}$ and the
$\theta\bar{\theta}\bar{\theta}$ terms (\ref{phistarphi}) is not a
chiral field. Following the method developed in \cite{miSUSY} we
decompose the $\star$-products of chiral fields into their
irreducible components by using the projectors defined in
\cite{wessbook}. The antichiral, chiral and transversal projectors
are defined as follows
\begin{eqnarray}
P_1 &=& \frac{1}{16} \frac{D^2 \bar{D}^2}{\Box}, \label{P1}\\
P_2 &=& \frac{1}{16} \frac{\bar{D}^2 D^2 }{\Box}, \label{P2}\\
P_T &=& -\frac{1}{8} \frac{D \bar{D}^2 D}{\Box}. \label{PT}
\end{eqnarray}

The chiral part of (\ref{phistarphi}) is undeformed and it is
given by (for details we refer to \cite{miSUSY})
\begin{eqnarray}
P_2(\Phi\star\Phi )&=&\Phi\Phi\nonumber\\
&=& A^2 + 2\sqrt{2}A\theta^\alpha\psi_\alpha
+ \theta\theta \Big( 2AH - \psi\psi\Big) \nonumber\\
&& + i\theta\sigma^m\bar{\theta}\Big( \p_m (A^2) \Big)  +
i\sqrt{2}\theta\theta\bar{\theta}_{\dot{\alpha}}\bar{\sigma}^{m\dot{\alpha}\alpha}
\big( \p_m(\psi_\alpha A)\big)\nn\\
&& +\frac{1}{4}\theta\theta\bar{\theta}\bar{\theta} \Box A^2 .\label{P2phistarphi}
\end{eqnarray}
The antichiral part reads
\begin{eqnarray}
P_1(\Phi\star\Phi)&=&- \frac{C^2}{2}H^2
-i\sqrt{2}C^2 H\bar{\theta}\bar\sigma^m\p_m \psi  + C^2\bar{\theta}\bar{\theta}\Big( - H\Box A
+ \frac{1}{2}(\p_m\psi)\sigma^m\bar{\sigma}^l(\p_l\psi) \big) \Big) \nonumber\\
&& + i\theta\sigma^m\bar{\theta}C^2H\p_m H  + \frac{\sqrt{2}}{2}\bar\theta\bar\theta C^2(-H\theta\Box  \psi+\theta\sigma^m\bar\sigma^n\p_n \psi\p_m H)\nonumber\\
&& -\frac{1}{8}\theta\theta\bar{\theta}\bar{\theta} C^2\Box H^2 .\label{P1phistarphi}
\end{eqnarray}
In this case there is no transverse part of $\Phi\star\Phi$,
\begin{equation}
P_T(\Phi\star\Phi)=0 . \label{PTphistarphi}
\end{equation}

Next, we calculate the $\star$-product of three chiral fields. The
following identity applies
\begin{eqnarray}
(\Phi\star\Phi)\star\Phi &=& (\Phi\cdot\Phi+P_1(\Phi\star\Phi))\star\Phi\nn\\
&=& \Phi\Phi\Phi-\frac{1}{32}C^2D^2(\Phi\Phi)D^2\Phi+P_1(\Phi\star\Phi)\Phi, \label{phistar3}
\end{eqnarray}
with
\begin{eqnarray}
-\frac{1}{32}C^2D^2(\Phi\Phi)D^2\Phi &=& C^2\Big[
-AH^2+\frac{1}{2} H(\psi\psi) - i\sqrt{2}AH\bar{\theta}\bar{\sigma}^n(\pa_n\psi)\nonumber\\ && -i\sqrt{2}(\bar{\theta}\bar{\sigma}^m\pa_m(A\psi))H
+\frac{i\sqrt2}{2}(\bar\theta\bar\sigma^n\pa_n\psi)
(\psi\psi)\nonumber\\
&& +\bar{\theta}\bar{\theta}\Big(-AH\Box A-\frac{1}{2}H\Box
A^2+\frac{1}{2}\psi\psi\Box A \nonumber\\
&& \hspace*{1cm} + \pa_m(A\psi)\sigma^m\bar\sigma^n(\pa_n\psi)\Big) \nonumber\\
&& +i\theta\sigma^m\bar\theta\pa_m\Big(AH^2-\frac12H\psi\psi\Big)\nonumber\\
&& +\frac{\sqrt2}{2}(\bar\theta\bar\theta)\Big(-AH\theta\Box\psi+\frac12(\theta\Box
\psi)(\psi\psi)-H\theta\Box(A\psi)\nonumber\\
&& +\theta\sigma^n\bar{\sigma}^m\pa_m(A\psi) (\pa_n H) +\frac{1}{2}\theta\sigma^l\bar{\sigma}^m(\pa_m\psi)\pa_l(2AH-\psi\psi)\Big)\nonumber\\
&& +\frac{1}{4}\theta\theta\bar{\theta}\bar{\theta}\Box\Big(-AH^2
+\frac{1}{2}(\psi\psi) H\Big) \Big] \label{form1}
\end{eqnarray}
and
\begin{eqnarray}
P_1(\Phi\star\Phi)\star\Phi &=& P_1(\Phi\star\Phi)\cdot\Phi\nn\\
&=& C^2\Big[-\frac{1}{2}AH^2-\frac{\sqrt2}{2}\theta\psi
H^2-\frac{1}{2}\theta\theta
H^3-i\sqrt2(\bar{\theta}\bar{\sigma}^m(\pa_m\psi)) AH\nn\\
&& +(\bar{\theta}\bar{\theta})(-HA\Box
A+\frac{1}{2}A(\pa_l\psi)\sigma^l\bar{\sigma}^m(\pa_m\psi))\nonumber\\
&& +i(\theta\sigma^l\bar{\theta})(\frac{1}{2}A(\pa_l H^2)
- \frac{1}{2}H^2(\pa_l A) + (\psi\sigma_l\bar{\sigma}^m(\pa_m\psi)) H)\nonumber\\
&& +i\sqrt{2}\Big(\frac{1}{4}(\theta\theta)(\bar{\theta}\bar{\sigma}^l\psi)
(\pa_l H^2) - \frac{5}{4}(\theta\theta)(\bar{\theta}\bar{\sigma}^m(\pa_m\psi))H^2\Big)\nn\\
&& +\frac{\sqrt2}{2} \bar{\theta}\bar{\theta}\Big( \theta\sigma^m\bar{\sigma}^n\pa_m(H\pa_n\psi)A - 2(\theta\psi)(H\Box
A - \frac{1}{2}(\pa_m\psi)\sigma^m\bar{\sigma}^n(\pa_n\psi))\nonumber\\
&& \hspace*{1.75cm} - (H\pa_l A)\theta\sigma^l\bar{\sigma}^m(\pa_m\psi) \Big)\nn\\
&& +\theta\theta\bar{\theta}\bar{\theta}\Big( -\frac{1}{8}A\Box
H^2 - \frac{9}{8}H^2\Box
A - \frac{1}{2}\psi\sigma^m\bar{\sigma}^n\pa_m(H\pa_n\psi)\nn\\
&& + H(\pa_m\psi)\sigma^m\bar{\sigma}^n(\pa_n\psi)
+ \frac{1}{4}(\pa_m A)(\pa^m H^2)\Big)
\Big]. \label{form2}
\end{eqnarray}
It is easy to see that
\begin{equation}
P_2(\Phi\star\Phi)\star\Phi=\Phi\Phi\Phi-\frac{1}{32}C^2D^2(\Phi\Phi)D^2\Phi . \label{form3}
\end{equation}
The projections are given by
\begin{eqnarray}
P_1(P_2(\Phi\star\Phi)\star\Phi)&=&-\frac{1}{32}C^2D^2(\Phi\Phi)D^2\Phi\nonumber\\
P_2(P_2(\Phi\star\Phi)\star\Phi)&=&\Phi\Phi\Phi . \label{form4}
\end{eqnarray}

\section{Invariants}

Let us now examine the transformation laws under the deformed SUSY
transformations (\ref{defsusytr}) of terms which could be relevant
for the construction of a SUSY invariant action.

There are two quadratic (in the number of fields)
invariants\footnote{Strictly speaking, terms $I_1$ and $I_2$ are
invariant only under the integral $\int{\mbox{d}}^4 x$, that is
when included in an action. Since the construction of an invariant
action is our aim, we continue with this abuse of notation and
call ''invariant`` all terms that under SUSY transformations
transform as total derivatives.}, $I_1$ and $I_2$:
\begin{eqnarray}
I_1 &=& P_2(\Phi\star\Phi)\Big|_{\theta\theta} = 2AH - \psi\psi , \label{I1}\\
I_2 &=& P_1(\Phi\star\Phi)\Big|_{\bar{\theta}\bar{\theta}} = -C^2\Big( H\Box
A - \frac{1}{2}(\p_m\psi)\sigma^m\bar{\sigma}^n(\p_n\psi)\Big) .\label{I2}
\end{eqnarray}
Their transformation laws are given by
\begin{eqnarray}
\delta^\star_\xi I_1  &=& 2i\sqrt{2}\bar{\xi}\bar{\sigma}^m
\p_m(A\psi) ,\label{deltaI1}\\
\delta^\star_\xi I_2 &=& \sqrt{2} C^2\xi\sigma^m\bar{\sigma}^n\p_m(H (\p_n\psi)). \label{deltaI2}
\end{eqnarray}

Looking at cubic terms we see that there are more candidates for
possible invariants. The first two are $I_3$ and $I_4$:
\begin{eqnarray}
I_3 &=& P_2(P_2(\Phi\star\Phi)\star\Phi)\Big|_{\theta\theta} = 3(A^2H - A\psi\psi), \label{I3}\\
I_4 &=&
P_1(P_2(\Phi\star\Phi)\star\Phi)\Big|_{\bar{\theta}\bar{\theta}}
\nonumber\\
&=& C^2\Big( -AH \Box A - \frac{1}{2}H\Box A^2 \nonumber\\
&& + \frac{1}{2}\psi\psi\Box A +
\p_m(A\psi)\sigma^m\bar{\sigma}^n(\p_n\psi)\Big) .\label{I4}
\end{eqnarray}
One can check that they indeed transform as total derivatives. Two
more candidates are given by
\begin{eqnarray}
I_5 &=& P_1(P_1(\Phi\star\Phi)\star\Phi)\Big|_{\bar\theta\bar\theta} = C^2\Big( -AH\Box
A + \frac{1}{2}A (\p_l\psi)\sigma^l\bar{\sigma}^m(\p_m\psi) \Big) ,\label{I5}\\
I_6 &=& P_2(P_1(\Phi\star\Phi)\star\Phi)\Big|_{\theta\theta} = -\frac{C^2}{2}H^3 .\label{I6}
\end{eqnarray}
However they do not transform as total derivatives
\begin{eqnarray}
\delta^\star_\xi I_5 &=& \frac{C^2}{2}\xi^\alpha\Big(
-2H( A\Box\psi_\alpha + \psi_\alpha\Box A)
+ 2(\sigma^m\bar{\sigma}^l)_\alpha^{\ \beta} (\p_l\psi_\beta)(\p_m H)A \nonumber\\
&& \hspace*{1.75cm} +\psi_\alpha(\p_l\psi)\sigma^l\bar{\sigma}^m(\p_m\psi)
\Big) \label{deltaI5}\\
&\neq& \p_m (\dots) ,\nonumber\\
\delta^\star_\xi I_6 &=& -\frac{3i}{\sqrt{2}}C^2 \bar{\xi}\bar{\sigma}^m(\p_m \psi)H^2 \label{deltaI6} \\
&\neq& \p_m(\dots) .\nonumber
\end{eqnarray}
Inclusion of these terms will not lead to a SUSY invariant action.
The last candidate for a cubic invariant is $I_7$:
\begin{eqnarray}
I_7 = P_2(P_1(\Phi\star\Phi)\star\Phi)\Big|_{\theta\theta\bar{\theta}\bar{\theta}} &=&
-\frac{C^2}{16}\Big( A\Box H^2  +5 H^2\Box A \label{I7}\\
&& - 4H(\p_m\psi)\sigma^m\bar{\sigma}^l(\p_l\psi) +
2\psi\sigma^m\bar{\sigma}^l\p_m(H(\p_l\psi))\Big) . \nonumber
\end{eqnarray}
Since we are interested in equations of motion we omitted a term
which is a total derivative in (\ref{I7}). Note also that the
terms
$P_1(P_1(\Phi\star\Phi)\star\Phi)\Big|_{\theta\theta\bar\theta\bar\theta}$
and
$P_2(P_1(\Phi\star\Phi)\star\Phi)\Big|_{\theta\theta\bar\theta\bar\theta}$
are equal up to a total derivative term and therefore lead to the
same equations of motion. Since $I_7$ is the highest component of
a superfield, under (\ref{defsusytr}) it transforms as a total
derivative and can be included in a SUSY invariant action.

\section{SUSY invariant Wess-Zumino model}

In order to write the SUSY invariant action we collect all
invariant terms and obtain the following Lagrangian
\begin{eqnarray}
{\cal L} &=&
\Phi^+\star\Phi\Big|_{\theta\theta\bar\theta\bar\theta} + \Big
[\frac{m}{2}\Big( P_2(\Phi\star\Phi)\Big|_{\theta\theta} +
aP_1(\Phi\star\Phi)
\Big|_{\bar\theta\bar\theta}\Big) \nn\\
&& +\frac{\lambda}{3}\Big( P_2(P_2(\Phi\star\Phi)\star\Phi)\Big|_{\theta\theta}
+bP_1(P_2(\Phi\star\Phi)\star\Phi)\Big|_{\bar\theta\bar\theta}\nn\\
&& + 2c(P_1+P_2)
(P_1(\Phi\star\Phi)\star\Phi)\Big|_{\theta\theta\bar{\theta}\bar{\theta}}
\Big) + {\mbox{ c.c. }} \Big],\label{Lagr}
\end{eqnarray}
with $m$, $\lambda$, $a$, $b$ and $c$ real constant parameters.
Terms
$P_1(P_1(\Phi\star\Phi)\star\Phi)\Big|_{\theta\theta\bar{\theta}\bar{\theta}}$
and
$P_2(P_1(\Phi\star\Phi)\star\Phi)\Big|_{\theta\theta\bar{\theta}\bar{\theta}}$
are equal up to a total derivative term and are therefore included
with the same coefficient. The action in component fields which
follows from (\ref{Lagr}) reads
\begin{eqnarray}
S &=& \int{\mbox{d}}^4 x\hspace{1mm} \Big\{
A^*\Box A+i\p_m\bar\psi\bar\sigma^m\psi+H^*H\nn\\
&& +m(AH-\frac12\psi\psi) + m(A^*H^*-\frac{1}{2}\bar\psi\bar\psi)\nonumber\\
&& +\lambda(A^2H-A\psi\psi) + \lambda\big( (A^*)^2H^*-A^*\bar\psi\bar\psi \big) \nn\\
&& + \Big[ C^2\Big( ma_1(\frac{1}{2}\psi\Box\psi-H\Box A)+\lambda a_2(-AH\Box A-\frac{1}{2}H(\Box A^2)\nn\\
&& +\frac{1}{2}\psi\psi(\Box A) + A\psi\Box\psi)
+\lambda a_3(-\frac{3}{2}H^2\Box A+\frac{3}{2}H(\p_m\psi)\sigma^m\bar{\sigma}^l(\p_l\psi)) \Big)\nn\\
&& + {\mbox{ c.c. }}\Big] \Big\}.
\label{Lagr.comp.f}
\end{eqnarray}
The coefficients $a$, $b$ and $c$ are related to $a_1$, $a_2$ and
$a_3$: $a/2 = a_1$, $b/3 = a_2$ and $c=a_3$. Note that
(\ref{Lagr.comp.f}) is the full action, i.e. no higher order terms
in the deformation parameter $C^{\alpha\beta}$ appear.

Varying the action (\ref{Lagr.comp.f}) with respect to the fields
$H$ and $H^*$ we obtain the equations of motion for these fields
\begin{eqnarray}
H^* &=& -mA - \lambda A^2 + ma_1C^2(\Box A) + \lambda a_2 C^2 \Big( A\Box A + \frac{1}{2}(\Box A^2)\Big) \nonumber\\
&& - \frac{3}{2}\lambda a_3 C^2\Big( -2H(\Box A) + (\p_m\psi)\sigma^m\bar{\sigma}^l(\p_l\psi) \Big) , \label{eomH*} \\
H &=& -mA^* - \lambda (A^*)^2 + ma_1\bar{C}^2(\Box A^*) + \lambda a_2 \bar{C}^2 \Big( A^*\Box A^* + \frac{1}{2}(\Box A^{*2})\Big) \nonumber\\
&& - \frac{3}{2}\lambda a_3 \bar{C}^2\Big( -2H^*(\Box A^*) + (\p_m\bar{\psi})\bar{\sigma}^m\sigma^l(\p_l\bar{\psi}) \Big) . \label{eomH}
\end{eqnarray}
Unlike in the undeformed theory, equations (\ref{eomH*}) and
(\ref{eomH}) are nonlinear in $H$ and $H^*$. Nevertheless they can
be solved
\begin{eqnarray}
H^* &=& \Big( 1 - 9(\lambda a_3)^2C^2\bar{C}^2 (\Box A^*)(\Box A) \Big)^{-1}\Big\{ -mA^* -\la (A^*)^2 \nonumber\\
&& + ma_1\bar{C}^2(\Box A^*) + \lambda a_2 \bar{C}^2 \Big( A^*\Box A^* + \frac{1}{2}(\Box A^*)^2\Big) \nonumber\\
&& -3\lambda a_3\bar{C}^2(\Box A^*)\big( mA + \lambda A^2\big)
- \frac{3}{2}\lambda a_3 \bar{C}^2(\p_m\bar{\psi})\bar{\sigma}^m\sigma^l(\p_l\bar{\psi}) \nonumber\\
&& + 3\lambda a_3\bar{C}^2(\Box A^*) \Big[ ma_1C^2(\Box A) + \lambda a_2 C^2 \Big( A\Box A + \frac{1}{2}(\Box A)^2\Big) \nonumber\\
&& - \frac{3}{2}\lambda a_3
C^2(\p_m\psi)\sigma^m\bar{\sigma}^l(\p_l\psi) \Big] \Big\},
\label{solH}
\end{eqnarray}
and similarly for $H$. These solutions we can expand up to second
order in the deformation parameter and insert in the action
(\ref{Lagr.comp.f}). The action then becomes
\begin{equation}
S = S_0 + S_2, \label{ScelobezF}
\end{equation}
with
\begin{eqnarray}
S_0 &=& \int{\mbox{d}}^4 x\hspace{1mm} \Big\{ A^*\Box A + i(\p_m\bar\psi)\bar{\sigma}^m\psi
-\frac{m}{2}\big( \psi\psi + \bar\psi\bar\psi \big) -\lambda \big( A^*\bar\psi\bar\psi + A\psi\psi \big) \nonumber\\
&& -m^2A^*A - m\lambda A(A^*)^2 - m\lambda A^*A^2 - \lambda^2A^2(A^*)^2 \Big\}
,\label{S0bezF} \\
S_2 &=& \int{\mbox{d}}^4 x\hspace{1mm} \Big\{
C^2ma_1\Big(\frac{1}{2}\psi(\Box\psi) + (\Box A)(mA^* + \lambda (A^*)^2) \Big) \nonumber\\
&& + C^2\lambda a_2\Big( \frac{1}{2}\psi\psi(\Box A) + A\psi(\Box\psi) + (mA^* + \lambda (A^*)^2)\big( A(\Box A) + \frac{1}{2}(\Box A^2) \big)\Big) \nonumber\\
&&- \frac{3}{2}C^2\lambda a_3(mA^* + \lambda (A^*)^2)\Big( (\Box A)(mA^* + \lambda (A^*)^2) + (\p_m\psi)\sigma^m\bar{\sigma}^l(\p_l\psi) \Big) \nonumber\\
&& +\bar{C}^2ma_1\Big(\frac{1}{2}\bar{\psi}(\Box\bar{\psi}) + (\Box A^*)(mA + \lambda A^2) \Big) \label{S2bezF}\\
&& + \bar{C}^2\lambda a_2\Big( \frac{1}{2}\bar{\psi}\bar{\psi}(\Box A^*) + A^*\bar{\psi}(\Box\bar{\psi}) + (mA + \lambda A^2)\big( A^*(\Box A^*) + \frac{1}{2}(\Box (A^*)^2) \big)\Big) \nonumber\\
&&- \frac{3}{2}\bar{C}^2\lambda a_3(mA + \lambda A^2)\Big( (\Box A^*)(mA + \lambda A^2) + (\p_m\bar{\psi})\bar{\sigma}^m\sigma^l(\p_l\bar{\psi}) \Big)
\Big\} . \nonumber
\end{eqnarray}

\section{Renormalizability properties: Two-point Green functions}

In this section we investigate some renormalizability properties
of our model. Using the background field method \cite{sw} and the
dimensional reduction\footnote{The method of dimensional
regularization has a draw-back that it might not preserve the
supersymmetry. Therefore one uses a modification of it, the
so-called dimensional reduction.} \cite{ggrs} the divergent part
of the effective action up to second order in fields is
calculated. Note that we work with the action (\ref{Lagr.comp.f})
and not with (\ref{ScelobezF}).

To start with, we rewrite the deformed action (\ref{Lagr.comp.f})
introducing the real fields $S$, $P$, $E$ and $G$ as
\begin{equation}
A= \frac{S+iP}{\sqrt 2}, \quad H
=\frac{E+iG}{\sqrt 2} \label{AH}
\end{equation}
and the Majorana spinor\footnote{The index M on the Majorana
spinors will be omitted in the following formulas.}
$\psi_M=\left(\matrix{\psi_\a\cr\bar\psi^{\dot \a}}\right)$. The
deformation parameter $C_{\alpha\beta}$ can be written in the
following way
\begin{eqnarray}
C_{\alpha\beta}= K_{ab}(\sigma^{ab}\varepsilon)_{\alpha\beta},
\quad \bar{C}_{\dot{\alpha}\dot{\beta}}=
K^*_{ab}(\varepsilon\bar{\sigma}^{ab})_{\dot{\alpha}\dot{\beta}}.
\label{n1}
\end{eqnarray}
Since $K_{ab}$ is a self dual tensor we write it as
\begin{equation}
K_{ab}=\kappa_{ab}+\frac{i}{2}\epsilon_{abcd}\kappa^{cd}, \label{Kab}
\end{equation}
where $\kappa_{ab}$ is a real antisymmetric tensor. In this way we
obtain
\begin{eqnarray}
C^2+\bar C^2 &=& 4\kappa_{ab}\kappa^{ab}\label{C+}\\
C^2-\bar C^2 &=& 2i\epsilon_{abcd}\kappa^{ab}\kappa^{cd} .
\label{C-}
\end{eqnarray}
In order to simplify our calculation we will assume that $C^2-\bar
C^2=0$. This choice can be obtained by setting $\kappa_{0i}=0$.

With all this and introducing $g=\frac{\lambda}{\sqrt 2}$ the
action (\ref{Lagr.comp.f}) becomes\footnote{In the notation of
\cite{wessbook} the matrix $\gamma_5$ and the Lorentz generators
$\Sigma^{mn}$ are defined as
\begin{equation}
\g^5=\g^0\g^1\g^2\g^3\ ,\quad \Sigma^{mn}=\frac14[\g^m,\g^n] .
\label{gamma5}
\end{equation}
}
$$S=S_0+S_2$$
with
\begin{eqnarray}
S_0 &=& \int{\mbox{d}}^4 x\hspace{1mm} \Big\{ \frac{1}{2} S\Box S+\frac12 P\Box
P-\frac12(i\bar\psi\g^m\pa_m\psi+m\bar\psi\psi)+\frac12(E^2+G^2)\nonumber\\
&& + m(SE-PG)-gS\bar\psi\psi+gP\bar\psi\g^5\psi\nonumber\\
&& + g(ES^2-EP^2-2SPG) \Big\}, \label{spL0}\\
S_2 &=& C^2\int{\mbox{d}}^4 x\hspace{1mm} \Big\{ ma_1(\frac12\bar\psi\Box\psi-E\Box S+G\Box P)\nonumber\\
&& + ga_2(PG\Box S-SE\Box S+PE\Box P+SG\Box P\nonumber\\
&& - \frac12(S^2\Box E-P^2\Box E-2SP\Box
G)+\frac{1}{2}\bar\psi\psi\Box
S\nonumber\\
&&-\frac{1}{2}\bar\psi\gamma^5\psi\Box P+\bar\psi\Box \psi
S-\bar\psi\gamma^5\Box \psi P)\nonumber\\
&& + \frac32 ga_3(-E^2\Box S+G^2\Box S+2EG\Box P-E\pa_m \bar\psi
\pa^m\psi\nonumber\\&+&G\pa_m\bar\psi\gamma^5\pa^m\psi-2\bar\psi\Sigma^{mn}\pa_n\psi\pa_m
E+2\bar\psi\Sigma^{mn}\g^5\pa_n\psi\pa_m G)\Big\}. \label{spL1}
\end{eqnarray}

We split the fields into their classical and quantum parts, for
example $E \to E + {\cal E}$. The action quadratic in quantum
fields is
\begin{eqnarray}
S^{(2)}=\frac{1}{2}\left(\matrix{\bar\Psi\ \cal{S}\
\cal{P}\ \cal{E}\ \cal{G} }\right)M\left(\matrix{\Psi\cr
\cal{S}\cr \cal{P}\cr \cal{E}\cr \cal{G} }\right)  , \label{S2kvpolja}
\end{eqnarray}
where $\Psi, \bar\Psi,\ \cal{S},\ \cal{P},\ \cal{E},\ \cal{G} $
are quantum fields. The one loop effective action is then
\begin{equation}
\Gamma
=\frac{i}{2}\Str\ln\Big[1+(\Box-m^2)^{-1}MC\Big]  ,\label{Gama}
\end{equation}
with
\begin{eqnarray}
C=\left(\matrix{-i\slash\pa+m&0&0&0&0\cr 0&1&0&-m&0&\cr
0&0&1&0&m\cr 0&-m&0&\Box& 0\cr 0&0&m&0&\Box}\right) . \label{C}
\end{eqnarray}
The matrix $MC$ can be decomposed into three parts
\begin{equation}
MC = N+T+V. \label{MC}
\end{equation}
The zeroth order (in the deformation parameter) term is given by
{\footnotesize
\begin{eqnarray}
N=2g\left( \matrix{(-S+\g^5P)(-i\slash \pa+m)&
-\psi&\gamma^5\psi&m\psi&m\gamma^5\psi\cr -\bar\psi(-i\slash
\pa+m)&(E-mS)&-(G+mP)&-mE+S\Box&-mG-P\Box\cr
\bar\psi\gamma^5(-i\slash \pa+m)&-G+mP&-E-mS&mG-P\Box&-mE-S\Box\cr
0&S&-P&-mS&-mP\cr 0&-P&-S&mP&-mS\cr }\right) . \label{N}
\end{eqnarray} }
The second order term (in the deformation parameter) which contains no fields is
\begin{eqnarray}
T= ma_1C^2\left(\matrix{\Box(-i\slash\pa+m)&0&0&0&0\cr
0&m\overleftarrow{\Box}&0&-\overleftarrow{\Box}\overrightarrow{\Box}&0&\cr
0&0&m\overleftarrow{\Box}&0&\overleftarrow{\Box}\overrightarrow{\Box}\cr
0&-\Box&0&m\Box& 0\cr 0&0&\Box&0&m\Box}\right) . \label{T}
\end{eqnarray}
The matrix $V$ is second order in the deformation parameter and
contains classical fields linearly. Its matrix elements are given
in Appendix A.

The one-loop divergent part of the effective action we calculate
up to second order in $g$, second order in fields (two-point
functions) and up to second order in the deformation parameter
$C_{\alpha\beta}$. Therefore the effective action is given by
\begin{eqnarray}
\Gamma &=& \frac{i}{2}\Str\ln\Big[1+(\Box-m^2)^{-1} (N+T+V)\Big]\nonumber\\
&=&\frac{i}{2}\Big[\Str((\Box-m^2)^{-1} (N+T+V))\nonumber\\
&&-\frac{1}{2}\Str((\Box-m^2)^{-1} N(\Box-m^2)^{-1}
N)\nonumber\\
&& -\Str((\Box-m^2)^{-1} N(\Box-m^2)^{-1} (T+V))\nonumber\\
&&+ \Str(((\Box-m^2)^{-1} N)^2(\Box-m^2)^{-1} T)\Big] .
\label{effS}
\end{eqnarray}
The calculation of divergent parts of supertraces is tedious but
straightforward and here we give only the results. The details are
given in Appendix B. Denoting $K=\Box-m^2$, we have
\begin{eqnarray}
\Str(K^{-1} (N+T+V)) &=& 0, \label{n+t+v} \\
\Str(K^{-1} NK^{-1} N) &=&
\frac{ig^2}{\pi^2\epsilon}\int{\mbox{d}}^4 x\hspace{1mm} \nonumber\\
&&\Big[S\Box S+P\Box P-\bar\psi i\slash\pa \psi+E^2+G^2\Big]
\label{nn}, \\
\Str\Big[K^{-1} N K^{-1} T\Big] &=& 0 ,\label{nt}\\
\Str(K^{-1} NK^{-1} V) &=&
-g^2C^2\frac{i}{2\pi^2\epsilon}\int{\mbox{d}}^4 x\hspace{1mm}
\Big[ 3a_3m^2(-2P\Box G\nonumber\\
&&-\bar\psi\Box \psi + 2S\Box E)\nonumber \\
&& +a_2((\Box S)^2+(\Box P)^2+4m^2S\Box S\nonumber\\
&&-\bar\psi i\slash\pa\Box \psi-2m^2\bar\psi i\slash
\pa\psi+4m^2P\Box P\nonumber\\&&+4m^2E^2+E\Box E+G\Box
G+4m^2G^2)\Big]  ,\label{nv}\\
\Str(K^{-1}NK^{-1}NK^{-1}T) &=&  \frac{2iC^2
a_1m^2g^2}{\pi^2\epsilon}\int{\mbox{d}}^4 x\hspace{1mm} \nonumber\\
&&\Big[S\Box
S+P\Box P-\bar\psi i\slash\pa \psi+E^2+G^2\Big] .\label{nnt}
\end{eqnarray}
In (\ref{nv}) terms $(\Box S)^2$ and $(\Box P)^2$ appear. Since
these terms do not have classical counterparts we take $a_2=0$.
Then the divergent part of the one loop  effective action
(\ref{effS}) is given by
\begin{eqnarray}
\Gamma_1 &=& \frac{g^2}{\pi^2\epsilon}\int{\mbox{d}}^4
x\hspace{1mm} \Big[\frac14(S\Box S+P\Box
P+\bar\psi i\slash\pa \psi+E^2+G^2)\nonumber\\
&& +\frac{3}{4}a_3C^2 m^2(2P\Box G + \bar\psi\Box \psi - 2S\Box
E)\nonumber\\
&& -C^2 a_1m^2(S\Box S+P\Box P-\bar\psi i\slash\pa
\psi+E^2+G^2)\Big] . \label{rezultat}
\end{eqnarray}

Let us now discuss the one-loop renormalizability properties of
our model. To cancel the divergences we have to add to the
classical Lagrangian the counterterms
\begin{equation}
{\cal L}_{B} = {\cal L}_0 + {\cal L}_2 - \Gamma_1 .\label{Lbare}
\end{equation}
In this way we obtain the bare Lagrangian ${\cal L}_B$.  It is
important to note that the term $I_7$ in the classical action
(\ref{Lagr.comp.f}) produces divergences proportional to $I_2$
(compare (\ref{nv}) and (\ref{spL1})), so both of them are
necessary in order to absorb the divergences in the effective
action. From the form of the bare Lagrangian we see that all
fields are renormalized in the same way:
\begin{equation}
S_0=\sqrt{Z}S,\>
P_0=\sqrt{Z}P,\> \psi_0=\sqrt{Z}\psi,\> E_0=\sqrt{Z}E,\> G_0=\sqrt{Z}G  , \label{barefields}
\end{equation}
with
\begin{equation}
Z = 1-\frac{g^2}{2\pi^2\epsilon}(1-4a_1m^2C^2) . \label{Z}
\end{equation}
The tadpole contributions add up to zero as in the commutative
case. Also, $\delta m=0$, i.e. there are no $\delta m
\bar\psi\psi$ and $\delta m(SE+PG)$ counterterms. It is obvious
that the deformation parameter has to be renormalized too,
\begin{equation}
C_0^2=\Big(1-\frac{3a_3g^2}{2\pi^2a_1\epsilon}\Big)C^2\
.\label{Cbare}
\end{equation}

The present analysis is not complete and we plan to consider the
vertex corrections in a forthcoming publication. From the vertex
corrections we should draw conclusions about the renormalization
of the coupling constant $g$ and about the renormalizability of
the full model.

Finally, let us make a comment concerning the non-renormalization
theorem. From (\ref{rezultat}) we see that the divergent part of
the effective action consists of the usual term
$(\Phi^+\Phi)\Big|_{\theta\theta\bar{\theta}\bar{\theta}} $ and a
new (compared to the undeformed case) term
$P_1(\Phi\star\Phi)\Big|_{\bar{\theta}\bar{\theta}}$. Both terms
are expressible as integrals over the whole superspace. In
particular, for the new term we have
\begin{eqnarray}
P_1(\Phi\star\Phi)\Big|_{\bar{\theta}\bar{\theta}} &=& \int{\mbox{d}}^4
x\hspace{1mm}{\mbox{d}}^2\hspace{1mm}\bar\theta{\mbox{d}}^2\hspace{1mm}\theta
\hspace{1mm}\theta\theta P_1(\Phi\star\Phi)\nonumber\\
&=& -\frac{1}{32}C^2\int{\mbox{d}}^4
x\hspace{1mm}{\mbox{d}}^2\hspace{1mm}\bar\theta{\mbox{d}}^2\hspace{1mm}\theta
\hspace{1mm}\theta\theta(D^2\Phi)(D^2\Phi)\nonumber\\
&=& \frac{1}{8}C^2\int{\mbox{d}}^4
x\hspace{1mm}{\mbox{d}}^2\hspace{1mm}\bar\theta{\mbox{d}}^2\hspace{1mm}\theta\hspace{1mm}\Phi(D^2\Phi) . \label{renom-thm}
\end{eqnarray}
We see that at the level of two-point Green functions there is no
need to deform the nonrenormalization theorem. This conclusion is
different from \cite{bfr}.

\section{Conclusions}

In order to see how a deformation by twist of the usual
Wess-Zumino model affects its renormalizability properties, we
considered a special example of the twist (\ref{twist}). Compared
with the undeformed SUSY Hopf algebra, the twisted SUSY Hopf
algebra is unchanged. In particular, the twisted coproduct is
undeformed, which leads to the undeformed Leibniz rule
(\ref{deftrlaw}). However, the notion of chirality is lost and we
have to apply the method of projectors introduced in
\cite{miSUSY}. By including all constructed invariants, we
formulate a deformation of the usual Wess-Zumino action
(\ref{Lagr.comp.f}). Finally, we discuss some preliminary
renormalizability properties of the model. As expected, there are
no tadpole diagrams and no mass renormalization counterterms. All
fields are renormalized in the same way, which is another property
of SUSY invariant theories. As the renormalization of the coupling
constant $g$ is concerned, at present we cannot say if it is
renormalized and how. However, we see that the freedom in choosing
terms in the action is partially fixed by demanding the
cancellation of divergences. That request leads to $a_2 =0$ and
additionally we see that both $a_1$ and $a_3$ terms are necessary.

Let us remark that the twist (\ref{twist}) leads to the
$\star$-product (\ref{star}) which has already been discussed in
\cite{Ferrara}. In that paper the deformed Wess-Zumino Lagrangian
has been constructed in two different ways. The difference was
present in the interaction terms. Namely, one can take the term
$\Phi^3_\star\Big|_{\theta\theta}$ which (since $\Phi^3_\star$ is
not chiral) breaks $1/2$ SUSY; this term is equal to our $I_6$
(\ref{I6}) and was not included in our deformed model
(\ref{Lagr.comp.f}) since it is not SUSY invariant. Adding its
complex conjugate breaks the full supersymmetry. The other
possibility which was considered in \cite{Ferrara} was to take the
term $\Phi^3_\star\Big|_{\theta\theta\bar{\theta}\bar{\theta}}$ as
an interaction term. Is is equal to our $I_7$ (\ref{I7}). Since it
is the highest component of the superfield $\Phi^3_\star$, it
transforms as a total derivative and the action is invariant under
the full supersymmetry. However, its commutative limit is zero and
it is not a deformation of the usual interaction term. The
commutative limit is obtained in \cite{Ferrara} by adding the term
$(\Phi^+)^3_\star$ which is undeformed and its complex conjugate
reproduces the proper commutative limit. We have seen that the
action with only the $I_7$ term is not renormalizable.

Renormalizability of the deformed Wess-Zumino models with the term
$\Phi^3_\star \propto H^3$ was studied, see for example
\cite{1/2WZrenorm}. To make these models renormalizable one has to
add additional terms to the original action. The main advantage of
our model is the absence of this problem. By including all
possible invariants from the beginning we see that no new terms
are needed to cancel the divergences that appear. However, our
results are not complete since we calculated here only the
divergences in the two-point functions. In the forthcoming paper
we will consider the vertex contributions and then we will be able
to tell if our present conclusions still hold.

\begin{appendix}

\section{Matrix elements of $V$}

The matrix elements of $V$ are given by
\begin{eqnarray}
V_{11}&=&gC^2\Big(a_2(\Box S+2S\Box-\gamma^5\Box
P-2P\gamma^5\Box)\nonumber\\
&& + 3a_3(E\Box+(\pa_m E)\pa^m-(\pa_m
G)\gamma^5\pa^m-G\gamma^5\Box\nonumber\\
&& - 2\Sigma^{mn}\pa_m E\pa_n+2\Sigma^{mn}\gamma^5\pa_mG\pa_n)\Big)(-i\slash
\pa+m) ,\label{V11}\\
V_{12}&=&gC^2\Big[a_2\Big(\psi\Box+2\Box \psi\Big)-3a_3m\Big(\Box
\psi+\pa_m\psi\pa^m-2\Sigma^{mn}\pa_n\psi\pa_m\Big)\Big] ,\label{V12}\\
V_{13} &=& gC^2\Big[a_2\Big(-\gamma^5\psi\Box-2\gamma^5\Box
\psi\Big) \nonumber\\
&& +3a_3m\Big(-\gamma^5\Box
\psi-\gamma^5\pa_m\psi\pa^m+2\Sigma^{mn}\g^5\pa_n\psi\pa_m\Big)\Big] ,\label{V13}\\
V_{14}&=&gC^2\Big[-ma_2\Big(\psi\Box+2\Box \psi\Big)+3a_3
\Big(\Box
\psi \nonumber\\
&& +\pa_m\psi\pa^m-2\Sigma^{mn}\pa_n\psi\pa_m\Big)\Box\Big] ,\label{V14}\\
V_{15}&=&gC^2\Big[ma_2\Big(-\gamma^5\psi\Box-2\gamma^5\Box
\psi\Big) \nonumber\\
&& +3a_3m\Big(-\gamma^5\Box
\psi-\gamma^5\pa_m\psi\pa^m+2\Sigma^{mn}\g^5\pa_n\psi\pa_m\Big)\Box\Big] ,\label{V15}\\
V_{21}&=&gC^2\Big[a_2\Big(\overleftarrow{\Box}\bar\psi+2\bar\psi\Box\Big)
(-i\slash \pa+m) ,\label{V21}\\
V_{22}&=&gC^2 \Big[-a_2(2E\Box+\Box E)+ma_2(\Box
S+S\Box+\overleftarrow{\Box}
S)+3ma_3\overleftarrow{\Box}E\Big] ,\label{V22}\\
V_{23}&=&gC^2\Big[a_2(\Box G+G\Box+\overleftarrow{\Box}
G) \nonumber\\
&& +ma_2(\Box P+P\Box+\overleftarrow{\Box}
P)+3ma_3\overleftarrow{\Box}G\Big] ,\label{V23}\\
V_{24}&=&gC^2\Big[ma_2(2E\Box+\Box E)-a_2(\Box
S+S\Box+\overleftarrow{\Box}
S)\Box-3a_3\overleftarrow{\Box}E\Box\Big] ,\label{V24}\\
V_{25}&=&gC^2\Big[ma_2(\Box G+G\Box+\overleftarrow{\Box}
G) \nonumber\\
&& +a_2(\Box P+P\Box+\overleftarrow{\Box}
P)\Box+3a_3\overleftarrow{\Box}G\Box\Big] ,\label{V25}\\
V_{31}&=&gC^2\Big[a_2
\Big(-\overleftarrow{\Box}\bar\psi\g^5-2\bar\psi\g^5\Box\Big)
(-i\slash \pa+m)\Big] ,\label{V31}\\
V_{32}&=&gC^2\Big[a_2(\Box G+G\Box+\overleftarrow{\Box}
G)-a_2(\Box P+P\Box+\overleftarrow{\Box}
P)-3ma_3\overleftarrow{\Box}G\Big] ,\label{V32}\\
V_{33}&=&gC^2 \Big[a_2(2E\Box+\Box E)+ma_2(\Box
S+S\Box+\overleftarrow{\Box}
S)+3ma_3\overleftarrow{\Box}E\Big] ,\label{V33}\\
V_{34} &=& gC^2\Big[-a_2m(\Box G+G\Box+\overleftarrow{\Box}
G) \nonumber\\
&& +a_2(\Box P+P\Box+\overleftarrow{\Box} P)\Box
+3a_3\overleftarrow{\Box}G\Box\Big]\ ,\label{V34}\\
V_{35}&=&gC^2\Big[ma_2(2E\Box+\Box E)+a_2(\Box
S+S\Box+\overleftarrow{\Box}
S)\Box+3a_3\overleftarrow{\Box}E\Box\Big] ,\label{V35}\\
V_{41}&=&gC^2\Big[3a_3\Big(-\pa^m\bar\psi\pa_m-2\pa_m\bar\psi\Sigma^{mn}\pa_n\Big)
(-i\slash
\pa+m)\Big] ,\label{V41}\\
V_{42}&=&gC^2\Big[-a_2(\Box S+S\Box+\overleftarrow{\Box} S)
+3a_3(m\Box S-E\Box)\Big] ,\label{V42}\\
V_{43}&=&gC^2\Big[a_2(\Box P+P\Box+\overleftarrow{\Box} P)
+3a_3(m\Box P+G\Box)\Big] ,\label{V43}\\
V_{44}&=&gC^2\Big[a_2gm(\Box S+S\Box+\overleftarrow{\Box} S)
+3a_3(-\Box S\Box+mE\Box)\Big] ,\label{V44}\\
V_{45}&=&gC^2\Big[ma_2(\Box P+P\Box+\overleftarrow{\Box}
P)+3a_3((\Box P)\Box+mG\Box)\Big] ,\label{V45}\\
V_{51}&=&gC^2\Big[3a_3\Big(\pa^m\bar\psi\gamma^5\pa_m-2\pa_m\bar\psi\Sigma^{mn}\pa_n\Big)(-i\slash \pa+m)\Big] , \label{V51}\\
V_{52}&=&gC^2\Big[a_2(\Box P+P\Box+\overleftarrow{\Box} P)
+3a_3(-m\Box P+G\Box)\Big] ,\label{V52}\\
V_{53}&=&gC^2\Big[a_2(\Box S+S\Box+\overleftarrow{\Box} S)
+3ma_3(m\Box S+E\Box)\Big] ,\label{V53}\\
V_{54}&=& -gC^2\Big[ma_2(\Box P+P\Box+\overleftarrow{\Box}
P)-3a_3(-mG\Box+(\Box P)\Box)\Big] ,\label{V54}\\
V_{55} &=& gC^2\Big[ma_2(\Box S+S\Box+\overleftarrow{\Box}
S)+3a_3(mE\Box+(\Box S))\Box\Big] .\label{V55}
\end{eqnarray}

\section{Calculation of Supertraces}

Here we calculate the divergent parts of  two supertraces:
$\Str(K^{-1} N K^{-1}V)$ and $\Str(K^{-1}NK^{-1}NK^{-1}T)$. The
following general formulas for the divergent parts of traces are
used
\begin{eqnarray}
\tr(K^{-1}fK^{-1}g)&=&\frac{i}{8\pi^2\epsilon}\int{\mbox{d}}^4
x\hspace{1mm} fg ,\label{f1}\\
\tr(\pa_nK^{-1}fK^{-1}g) &=&\frac{i}{16\pi^2\epsilon}\int{\mbox{d}}^4
x\hspace{1mm} \pa_nfg ,\label{f2}\\
\tr(\pa_nK^{-1}f\pa_mK^{-1}g) &=& -\frac{i}{8\pi^2\epsilon}\int{\mbox{d}}^4
x\hspace{1mm} \label{f3}\\
&&\Big(\frac{1}{6} \pa_m\pa_nfg+\frac{1}{12}\eta_{mn}\Box
f g-\frac{1}{2}\eta_{mn}m^2fg\Big) , \nonumber\\
\tr(K^{-1}f\pa_aK^{-1}g\pa_bK^{-1}h) &=& \frac{i}{32\pi^2\epsilon}\eta_{ab}\int{\mbox{d}}^4 x\hspace{1mm}
fgh ,\label{f4}\\
\tr(K^{-1}f) &=& \frac{i}{8\pi^2\epsilon}m^2\int{\mbox{d}}^4
x\hspace{1mm} f ,\label{f5} \\
\tr(\pa_aK^{-1}f) &=&0 , \label{f6}\\
\tr(\Box K^{-1}f) &=& \frac{im^4}{8\pi^2\epsilon}\int{\mbox{d}}^4 x\hspace{1mm}
f ,\label{f7}\\
\tr(\Box^2K^{-1}f) &=& \frac{im^6}{16\pi^2\epsilon}\int{\mbox{d}}^4
x\hspace{1mm} f  . \label{f8}
\end{eqnarray}

\begin{itemize}
\item $\Str(K^{-1} N K^{-1}V)$
\vspace*{0.3cm}

Using the definition of Supertrace we obtain
\begin{eqnarray}
\Str(K^{-1} NK^{-1} V) &=& -\sum_i\tr(K^{-1} N_{1i}K^{-1}
V_{i1})\nonumber\\
&& +\sum_i\tr(K^{-1} N_{2i}K^{-1}
V_{i2}) + \dots \nonumber\\
&& + \sum_i\tr(K^{-1} N_{5i}K^{-1}
V_{i5}).\label{nvracun}
\end{eqnarray}
The terms in (\ref{nvracun}) are
\begin{eqnarray}
&&\hspace*{-0.5cm}\tr\Big[K^{-1} N_{11}K^{-1}
V_{11}\Big] \nonumber\\
&&= g^2C^2\frac{i}{2\pi^2\epsilon}\int{\mbox{d}}^4
x\hspace{1mm} \Big[ a_2((\Box S)^2 + (\Box
P)^2 - 4m^2S\Box S - 20m^4S^2 - 4m^4P^2)\nonumber\\
&&\hspace*{0.5cm} + 3a_3( -m^2 P\Box G - 10m^4SE + m^2 S\Box E - 2m^4PG)
\Big] , \label{NV1}\\
&&\hspace*{-0.5cm}\tr\Big[K^{-1}N_{23}K^{-1}V_{32}+K^{-1}N_{32}K^{-1}V_{23}\Big]\nonumber\\
&&=\frac{ig^2C^2}{4\pi^2\epsilon}\int{\mbox{d}}^4 x\hspace{1mm}
\Big[-2 a_2(G\Box G+4m^2G^2) \nonumber\\
&& \hspace*{0.5cm}+ 2 m^2a_2(P\Box P+4m^2P^2)+12a_3m^4PG \Big] , \label{NV3}\\
&&\hspace*{-0.5cm}\tr\Big[K^{-1}N_{45}K^{-1}V_{54}+K^{-1}N_{54}K^{-1}V_{45}\Big] \nonumber\\
&&=\frac{i}{2\pi^2\epsilon}mg^2C^2\int{\mbox{d}}^4 x\hspace{1mm}
\Big[ a_2m(P\Box P+4m^2P^2) + 6a_3m^3PG \Big] , \label{NV4}\\
&&\hspace*{-0.5cm}\tr\Big[K^{-1}V_{41}K^{-1}N_{14}+K^{-1}V_{51}K^{-1}N_{15}\Big] \nonumber\\
&&= 3g^2C^2 a_3m^2\frac{i}{4\pi^2\epsilon}\int{\mbox{d}}^4
x\hspace{1mm} \bar\psi\Box \psi  ,\label{NV7}\\
&&\hspace*{-0.5cm}\tr\Big[K^{-1}N_{22}K^{-1}V_{22} + K^{-1}N_{33}K^{-1}V_{33}\Big] \nonumber\\
&&= g^2C^2\frac{i}{2\pi^2\epsilon}
\int{\mbox{d}}^4 x\hspace{1mm} \Big[ -2a_2(2m^2E^2+\frac{1}{2}E\Box E) \nonumber\\
&&\hspace*{0.5cm}-m^2a_2(S\Box S+4m^2S^2)-6a_3m^4SE\Big] ,\label{NV8}\\
&&\hspace*{-0.5cm}\tr\Big[K^{-1}N_{44}K^{-1}V_{44} + K^{-1}N_{55}K^{-1}V_{55}\Big]\nonumber\\
&& = -g^2m^2C^2\frac{i}{2\pi^2\epsilon}\int{\mbox{d}}^4 x\hspace{1mm} \Big[ a_2(4m^2S^2+S\Box S)+6a_3m^2SE\Big]. \label{NV9}\\
&&\hspace*{-0.5cm}\tr\Big[K^{-1}N_{42}K^{-1}V_{24} +
K^{-1}N_{24}K^{-1}V_{42} + K^{-1}N_{35}K^{-1}V_{53}+K^{-1}N_{53}K^{-1}V_{35}\Big]\nonumber\\
&& =\frac{3ig^2C^2}{2\pi^2\epsilon}\int{\mbox{d}}^4 x\hspace{1mm}
\Big[-2a_2(m^2S\Box S + 2m^4S^2)\nonumber\\
&& \hspace*{0.5cm}- m^2a_3( 6m^2 SE + E\Box S) \Big], \label{NV2}  \\
&&\hspace*{-0.5cm}\tr\Big[K^{-1}N_{25}K^{-1}V_{52} + K^{-1}N_{52}K^{-1}V_{25} + K^{-1}N_{34}K^{-1}V_{43} + K^{-1}N_{43}K^{-1}V_{34}\Big]\nonumber\\
&& =\frac{3i}{2\pi^2\epsilon}g^2C^2\int{\mbox{d}}^4 x\hspace{1mm}
\Big[-2a_2(m^2P\Box P + 2m^4P^2)\nonumber\\
&& \hspace*{0.5cm}- m^2 a_3( 6m^2PG - G\Box P) \Big]  ,\label{NV5}\\
&&\hspace*{-0.5cm}\tr\Big[K^{-1}N_{21}K^{-1}V_{12} - K^{-1}N_{12}K^{-1}V_{21} + K^{-1}N_{31}K^{-1}V_{13} - K^{-1}N_{13}K^{-1}V_{31}\Big]\nonumber\\
&& = g^2C^2\frac{i}{2\pi^2\epsilon}\int{\mbox{d}}^4 x\hspace{1mm} \Big[a_2(2im^2\bar\psi\slash\pa\psi +i\bar\psi\slash\pa\Box\psi)\nonumber\\
&&
\hspace*{0.5cm}+\frac{3}{2}a_3m^2\bar\psi\Box\psi\Big].\label{NV6}
\end{eqnarray}
Adding all the terms (\ref{NV1})-(\ref{NV9}) we obtain
\begin{eqnarray}
\Str(K^{-1} NK^{-1} V) &=& - g^2C^2\frac{i}{2\pi^2\epsilon}\int{\mbox{d}}^4
x\hspace{1mm} \Big[ 3a_3m^2(-2P\Box G\nonumber\\
&& -\bar\psi\Box \psi + 2S\Box E)\nonumber\\
&& + a_2((\Box S)^2+(\Box P)^2+4m^2S\Box S\nonumber\\
&& -\bar\psi i\slash\pa\Box \psi-2m^2\bar\psi i\slash \pa\psi+4m^2P\Box
P\nonumber\\
&& +4m^2E^2+E\Box E+G\Box G+4m^2G^2)\Big] .\label{NVsum}
\end{eqnarray}

\item $\Str(K^{-1}NK^{-1}NK^{-1}T)$
\vspace*{0.3cm}

Again, from the definition of Supertrace it follows
\begin{eqnarray}
\Str(K^{-1}NK^{-1}NK^{-1}T) &=& -\tr(K^{-1}N_{1i}K^{-1}N_{ij}K^{-1}T_{j1})\nonumber\\
&& + \tr(K^{-1}N_{2i}K^{-1}N_{ij}K^{-1}T_{j2}) \nonumber\\
&& + \dots \nonumber\\
&& + \tr(K^{-1}N_{5i}K^{-1}N_{ij}K^{-1}T_{j5}) .\label{NNT}
\end{eqnarray}
The divergences appearing in (\ref{NNT}) are
\begin{eqnarray}
&&\hspace*{-0.5cm}\tr(K^{-1}N_{11}K^{-1}N_{11}\Box (-i\slash\pa+m))\nonumber\\
&& =4 g^2m\frac{i}{8\pi^2\epsilon}\int{\mbox{d}}^4 x\hspace{1mm}
\Big[-4S\Box S\nonumber\\
&& \hspace*{0.5cm}-4P\Box P+40m^2S^2+8m^2P^2\Big] , \label{NNT1}\\
&&\hspace*{-0.5cm}\tr(K^{-1}N_{22}K^{-1}N_{24}K^{-1}\Box) \nonumber\\
&& =\frac{4g^2i}{8\pi^2\epsilon}\int{\mbox{d}}^4 x\hspace{1mm} \Big[-mE^2+4m^2SE-3m^3S^2\Big], \label{NNT2}\\
&&\hspace*{-0.5cm}\tr(K^{-1}N_{23}K^{-1}N_{34}K^{-1}\Box)\nonumber\\
&&= \frac{4g^2i}{8\pi^2\epsilon}\int{\mbox{d}}^4 x\hspace{1mm}
\Big[-mG^2+2mGP+3m^2P^2\Big] ,\label{NNT3}\\
&&\hspace*{-0.5cm}\tr(K^{-1}N_{24}K^{-1}N_{44}K^{-1}\Box)\nonumber\\
&& = \frac{4g^2i}{8\pi^2\epsilon}\int{\mbox{d}}^4 x\hspace{1mm}
\Big[m^2SE-3m^3S^2\Big] ,\label{NNT4}\\
&&\hspace*{-0.5cm}\tr(K^{-1}N_{25}K^{-1}N_{54}K^{-1}\Box)\nonumber\\
&&=-4g^2\frac{i}{8\pi^2\epsilon}\int{\mbox{d}}^4 x\hspace{1mm}
\Big[3m^3P^2+m^2PG\Big] ,\label{NNT5}\\
&&\hspace*{-0.5cm}\tr(K^{-1}N_{32}K^{-1}N_{25}K^{-1}\Box)\nonumber\\
&& =\frac{4g^2i}{8\pi^2\epsilon}\int{\mbox{d}}^4 x\hspace{1mm}
\Big[2m^2PG+m G^2-3m^3P^2\Big] ,\label{NNT6}\\
&&\hspace*{-0.5cm}\tr(K^{-1}N_{33}K^{-1}N_{35}K^{-1}\Box)\nonumber\\
&& = 4g^2\frac{i}{8\pi^2\epsilon}\int{\mbox{d}}^4 x\hspace{1mm}
\Big[4m^2ES+3m^3S^2+mE^2\Big] ,\label{NNT7}\\
&&\hspace*{-0.5cm}\tr(K^{-1}N_{34}K^{-1}N_{45}K^{-1}\Box)\nonumber\\
&&=4g^2 \frac{i}{8\pi^2\epsilon}\int{\mbox{d}}^4 x\hspace{1mm}
\Big[3m^3P^2-m^2GP\Big], \label{NNT8}\\
&&\hspace*{-0.5cm}\tr(K^{-1}N_{35}K^{-1}N_{55}K^{-1}\Box) \nonumber\\
&&=4g^2 \frac{i}{8\pi^2\epsilon}\int{\mbox{d}}^4 x\hspace{1mm}
\Big[m^2ES+3m^3S^2\Big], \label{NNT9}\\
&&\hspace*{-0.5cm}m\tr(K^{-1}N_{24}K^{-1}N_{42}K^{-1}\Box) \nonumber\\
&&=4g^2 \frac{i}{8\pi^2\epsilon}\int{\mbox{d}}^4 x\hspace{1mm}
\Big[3m^3S^2-m^2ES\Big] ,\label{NNT10}\\
&&\hspace*{-0.5cm}m\tr(K^{-1}N_{25}K^{-1}N_{52}K^{-1}\Box)\nonumber\\
&&=4g^2\frac{i}{8\pi^2\epsilon}\int{\mbox{d}}^4 x\hspace{1mm}
\Big[3m^3P^2+m^2PG\Big], \label{NNT11}\\
&&\hspace*{-0.5cm}m\tr(K^{-1}N_{35}K^{-1}N_{53}K^{-1}\Box)\nonumber\\
&&=4g^2\frac{i}{8\pi^2\epsilon}\int{\mbox{d}}^4 x\hspace{1mm} \Big[
m^2ES+3m^3S^2\Big], \label{NNT12}\\
&&\hspace*{-0.5cm}m\tr(K^{-1}N_{34}K^{-1}N_{43}K^{-1}\Box)\nonumber\\
&&=4g^2 \frac{i}{8\pi^2\epsilon}\int{\mbox{d}}^4 x\hspace{1mm} \Big[-
m^2GP+3m^3P^2\Big], \label{NNT13}\\
&&\hspace*{-0.5cm}\tr(K^{-1}N_{33}K^{-1}N_{33}K^{-1}\Box) \nonumber\\
&&=4g^2\frac{i}{8\pi^2\epsilon}\int{\mbox{d}}^4 x\hspace{1mm} (
E+mS)^2 ,\label{NNT14}\\
&&\hspace*{-0.5cm}\tr(K^{-1}N_{32}K^{-1}N_{23}K^{-1}\Box)\nonumber\\
&&=4g^2\frac{i}{8\pi^2\epsilon}\int{\mbox{d}}^4 x\hspace{1mm} \Big[
mG^2-m^3P^2\Big], \label{NNT15}\\
&&\hspace*{-0.5cm}m\tr(K^{-1}N_{35}K^{-1}N_{53}K^{-1}\Box)\nonumber\\
&&=4g^2\frac{i}{8\pi^2\epsilon}\int{\mbox{d}}^4 x\hspace{1mm} \Big[
m^2ES+3m^3S^2\Big], \label{NNT16}\\
&&\hspace*{-0.5cm}\tr(K^{-1}N_{54}K^{-1}N_{45}K^{-1}\Box)\nonumber\\
&&=-4g^2\frac{i}{8\pi^2\epsilon}\int{\mbox{d}}^4 x\hspace{1mm} m^3P^2, \label{NNT17}\\
&&\hspace*{-0.5cm}\tr(K^{-1}N_{55}K^{-1}N_{55}K^{-1}\Box)\nonumber\\
&&=4g^2\frac{i}{8\pi^2\epsilon}\int{\mbox{d}}^4 x\hspace{1mm} m^3S^2, \label{NNT18}\\
&&\hspace*{-0.5cm}\tr(K^{-1}N_{22}K^{-1}N_{22}K^{-1}\Box)\nonumber\\
&&=4g^2\frac{i}{8\pi^2\epsilon}\int{\mbox{d}}^4 x\hspace{1mm}
(E-mS)^2, \label{NNT19}\\
&&\hspace*{-0.5cm}\tr(K^{-1}N_{21}K^{-1}N_{14}K^{-1}\Box)\nonumber\\
&&=4g^2\frac{im}{8\pi^2\epsilon}\int{\mbox{d}}^4 x\hspace{1mm} \Big[
\frac{i}{2}\bar\psi\slash\pa\psi-m\bar\psi\psi\Big], \label{NNT20}\\
&&\hspace*{-0.5cm}\tr(K^{-1}N_{31}K^{-1}N_{15}K^{-1}\Box)\nonumber\\
&&=-4g^2\frac{im}{8\pi^2\epsilon}\int{\mbox{d}}^4 x\hspace{1mm}
\Big[
\frac{i}{2}\bar\psi\slash\pa\psi+m\bar\psi\psi\Big], \label{NNT21}\\
&&\hspace*{-0.5cm}\tr(K^{-1}N_{31}K^{-1}N_{13}K^{-1}\Box)\nonumber\\
&&=-4mg^2\frac{i}{8\pi^2\epsilon}\int{\mbox{d}}^4 x\hspace{1mm} \Big[
\frac{i}{2}\bar\psi\slash\pa\psi+m\bar\psi\psi\Big], \label{NNT22}\\
&&\hspace*{-0.5cm}\tr(K^{-1}N_{21}K^{-1}N_{12}K^{-1}\Box)\nonumber\\
&&=4g^2m\frac{i}{8\pi^2\epsilon}\int{\mbox{d}}^4 x\hspace{1mm} \Big[
-\frac{i}{2}\bar\psi\slash\pa\psi+m\bar\psi\psi\Big], \label{NNT23}\\
&&\hspace*{-0.5cm}\tr(K^{-1}N_{13}K^{-1}N_{31}K^{-1}(-i\slash
\pa+m)\Box)+\tr(K^{-1}N_{12}K^{-1}N_{21}K^{-1}(-i\slash
\pa+m)\Box)\nonumber\\
&&= 8g^2\frac{i}{8\pi^2\epsilon}\int{\mbox{d}}^4 x\hspace{1mm} i\bar\psi\slash\pa\psi, \label{NNT24}\\
&&\hspace*{-0.5cm}\tr(K^{-1}N_{42}K^{-1}N_{22}\Box K^{-1}\Box)\nonumber\\
&&=4g^2\frac{i}{8\pi^2\epsilon}\int{\mbox{d}}^4 x\hspace{1mm}
\Big[3m^2SE-3m^3S^2\Big], \label{NNT25}\\
&&\hspace*{-0.5cm}\tr(K^{-1}N_{43}K^{-1}N_{32}\Box K^{-1}\Box)\nonumber\\
&&=4g^2\frac{i}{8\pi^2\epsilon}\int{\mbox{d}}^4 x\hspace{1mm}
\Big[3m^2PG-3m^3P^2\Big], \label{NNT26}\\
&&\hspace*{-0.5cm}\tr(K^{-1}N_{44}K^{-1}N_{42}\Box K^{-1}\Box)\nonumber\\
&&=-12g^2 \frac{i}{8\pi^2\epsilon}\int{\mbox{d}}^4 x\hspace{1mm} m^3S^2, \label{NNT27}\\
&&\hspace*{-0.5cm}\tr(K^{-1}N_{45}K^{-1}N_{52}\Box K^{-1}\Box)\nonumber\\
&&=12g^2\frac{i}{8\pi^2\epsilon}\int{\mbox{d}}^4 x\hspace{1mm} m^3P^2, \label{NNT28}\\
&&\hspace*{-0.5cm}\tr(K^{-1}N_{54}K^{-1}N_{43}\Box K^{-1}\Box)\nonumber\\
&&=-12g^2 \frac{i}{8\pi^2\epsilon}\int{\mbox{d}}^4 x\hspace{1mm} m^3P^2, \label{NNT29}\\
&&\hspace*{-0.5cm}\tr(K^{-1}N_{55}K^{-1}N_{53}\Box K^{-1}\Box)\nonumber\\
&&=12g^2\frac{i}{8\pi^2\epsilon}\int{\mbox{d}}^4 x\hspace{1mm} m^3S^2, \label{NNT30}\\
&&\hspace*{-0.5cm}\tr(K^{-1}N_{52}K^{-1}N_{23}\Box K^{-1}\Box)\nonumber\\
&&=12g^2\frac{i}{8\pi^2\epsilon}\int{\mbox{d}}^4 x\hspace{1mm}
\Big[m^2PG+m^3P^2\Big], \label{NNT31}\\
&&\hspace*{-0.5cm}\tr(K^{-1}N_{53}K^{-1}N_{33}\Box K^{-1}\Box)\nonumber\\
&&=12g^2\frac{i}{8\pi^2\epsilon}\int{\mbox{d}}^4 x\hspace{1mm}
\Big[m^2SE+m^3S^2\Big]. \label{NNT33}
\end{eqnarray}

Summing the terms (\ref{NNT1})-(\ref{NNT33}) we obtain
\begin{eqnarray}
\Str(K^{-1}NK^{-1}NK^{-1}T)&=& \frac{2ia_1C^2
m^2g^2}{\pi^2\epsilon} \label{NNTsum}\\
&&\int{\mbox{d}}^4 x\hspace{1mm}
\Big[S\Box S+P\Box P-\bar\psi i\slash\pa
\psi +E^2+G^2\Big] .\nonumber
\end{eqnarray}

\end{itemize}

\end{appendix}

\vspace*{0.5cm}
\begin{flushleft}
{\Large {\bf Acknowledgments}}
\end{flushleft}

The work of M.D. and V.R. is supported by the
project $141036$ of the Serbian Ministry of Science. M.D. also thanks
INFN Gruppo collegato di Alessandria for their financial support
during one year stay in Alessandria, Italy where a part of this
work was completed.

\end{document}